\documentstyle[epsf]{mn}
\newif\ifAMStwofonts
\ifoldfss
  \ifCUPmtlplainloaded \else
    \NewTextAlphabet{textbfit} {cmbxti10} {}
    \NewTextAlphabet{textbfss} {cmssbx10} {}
    \NewMathAlphabet{mathbfit} {cmbxti10} {} % for math mode
    \NewMathAlphabet{mathbfss} {cmssbx10} {} %  "   "    "
  \fi
  \ifAMStwofonts
    \ifCUPmtlplainloaded \else
      \NewSymbolFont{upmath} {eurm10}
      \NewSymbolFont{AMSa} {msam10}
      \NewMathSymbol{\upi}     {0}{upmath}{19}
      \NewMathSymbol{\umu}     {0}{upmath}{16}
      \NewMathSymbol{\upartial}{0}{upmath}{40}
      \NewMathSymbol{\leqslant}{3}{AMSa}{36}
      \NewMathSymbol{\geqslant}{3}{AMSa}{3E}

    \fi
  \fi
\fi % End of OFSS

\ifnfssone
  \newmathalphabet{\mathit}
  \addtoversion{normal}{\mathit}{cmr}{m}{it}
  \addtoversion{bold}{\mathit}{cmr}{bx}{it}
  \newmathalphabet{\mathbfit} % math mode version of \textbfit{..}
  \addtoversion{normal}{\mathbfit}{cmr}{bx}{it}
  \addtoversion{bold}{\mathbfit}{cmr}{bx}{it}
  \newmathalphabet{\mathbfss} % math mode version of \textbfss{..}
  \addtoversion{normal}{\mathbfss}{cmss}{bx}{n}
  \addtoversion{bold}{\mathbfss}{cmss}{bx}{n}
  \ifAMStwofonts
    \ifCUPmtlplainloaded \else
      \UseAMStwoboldmath
      \makeatletter
      \new@mathgroup\upmath@group
      \define@mathgroup\mv@normal\upmath@group{eur}{m}{n}
      \define@mathgroup\mv@bold\upmath@group{eur}{b}{n}
      \edef\UPM{\hexnumber\upmath@group}
      \new@mathgroup\amsa@group
      \define@mathgroup\mv@normal\amsa@group{msa}{m}{n}
      \define@mathgroup\mv@bold\amsa@group{msa}{m}{n}
      \edef\AMSa{\hexnumber\amsa@group}
      \makeatother
      \mathchardef\upi="0\UPM19
      \mathchardef\umu="0\UPM16
      \mathchardef\upartial="0\UPM40
      \mathchardef\leqslant="3\AMSa36
      \mathchardef\geqslant="3\AMSa3E
    \fi
  \fi
\fi % End of NFSS release 1

\ifnfsstwo
  \DeclareMathAlphabet{\mathbfit}{OT1}{cmr}{bx}{it}
  \SetMathAlphabet\mathbfit{bold}{OT1}{cmr}{bx}{it}
  \DeclareMathAlphabet{\mathbfss}{OT1}{cmss}{bx}{n}
  \SetMathAlphabet\mathbfss{bold}{OT1}{cmss}{bx}{n}
  \ifAMStwofonts
    \ifCUPmtlplainloaded \else
      \DeclareSymbolFont{UPM}{U}{eur}{m}{n}
      \SetSymbolFont{UPM}{bold}{U}{eur}{b}{n}
      \DeclareSymbolFont{AMSa}{U}{msa}{m}{n}
      \DeclareMathSymbol{\upi}{0}{UPM}{"19}
      \DeclareMathSymbol{\umu}{0}{UPM}{"16}
      \DeclareMathSymbol{\upartial}{0}{UPM}{"40}
      \DeclareMathSymbol{\leqslant}{3}{AMSa}{"36}
      \DeclareMathSymbol{\geqslant}{3}{AMSa}{"3E}
    \fi
  \fi
\fi % End of NFSS release 2

\ifCUPmtlplainloaded \else
  \ifAMStwofonts \else % If no AMS fonts
    \def\upi{\pi}
    \def\umu{\mu}
    \def\upartial{\partial}
  \fi
\fi

\title[HAR observations of WR~112]
{Detection of a sub-arcsecond dust shell around the Wolf-Rayet star
WR~112
\thanks{Based on observations collected
at TIRGO (Gornergrat, Switzerland) and at Calar Alto (Spain).
TIRGO is operated by CNR -- CAISMI Arcetri,
Italy. Calar Alto is operated by the German-Spanish Astronomical
Center.}
}
\author[Ragland and Richichi]
	{S. Ragland and A. Richichi\\
Osservatorio Astrofisico di Arcetri,
L.go Enrico Fermi 5, I-50125 Firenze, Italy\\
sam@arcetri.astro.it; richichi@arcetri.astro.it
}
\date{Received/Accepted}

\pagerange{\pageref{firstpage}--\pageref{lastpage}}
\pubyear{1994}

\begin{document}

\maketitle

\label{firstpage}

\begin{abstract}
A lunar occultation event of the Wolf-Rayet star 
WR~112 (type WC9) 
has been observed simultaneously from two independent telescopes at
$\lambda$ = 2.2$\mu$m, allowing us to investigate this source with an angular 
resolution of $\approx$ 0.003 arc-seconds. We have detected 
a circumstellar dust envelope whose brightness distribution can be approximately 
fitted by
a gaussian with a  
FWHM of $\approx$0.06 arc-seconds ($\approx$ 10$^{15}$cm). 
We present and discuss the reconstructed brightness profile,
which shows an asymmetry in the
radial dust distribution. 
The derived dust grain 
temperature at the inner dust zone of $\approx$1150\,K
%suggests that the shell
%is situated approximately at the expected position of the condensation zone and
is consistent with available model calculations.
There is no signature of the
central star from our observations, providing a direct 
confirmation that the
circumstellar shell emission dominates over the photospheric 
emission at 2.2$\mu$m as predicted by fits to the spectral energy distribution. 
%Also, we do not detect the bright rim predicted by
%model calculations, but this is not conclusive as the
%1--D integration intrinsic in lunar occultation observations
%effectively does not permit  such a detection.
%%expected bright rim will not be apparent in the 1--D 
%%brightness profiles recovered
%%from our observations.
Further lunar occultation observations at different position angles are
essential to reconstruct the 2--D image of the dust shell around WR~112.
The current series of 
lunar occultations of WR~112 will continue to the end of 1999 and will 
be visible for all equatorial and southern latitude observatories.
\end{abstract}

\begin{keywords}
occultations -- stars: fundamental parameters --
stars: Wolf-Rayet  --
stars: circumstellar matter --
infrared: stars
\end{keywords}

\section{Introduction}\label{introduction}
Wolf-Rayet (WR) stars represent  a small population of 
evolved
high-mass stars. They are characterized by strong, broad emission line
features in their optical spectra, indicative of 
powerful stellar winds and in turn of 
mass loss (Abbott \& Conti 1987). This latter is inferred
also from P~Cygni profiles in the UV and visual, from free-free emissions 
and thermal
dust emissions in the infrared,
from continuum excess radiation in radio, and
from H$_\alpha$ 
and forbidden--line imaging in the optical.
The total wind energy emitted by a WR
star during its lifetime is comparable to the energy of a supernova
(van der Hucht, Williams \& Th\'{e} 1987).

%WR stars are divided into two main classes, namely, WN (nitrogen
%sequence) and
%WC (carbon sequence) type stars. An another class, WO (oxygen sequence)
%can also be seen in the literature 
%(Barlow \& Hummer, 1982).
%The fundamental parameters of WR stars such as
%luminosity and effective temperature are highly controversial
%given the complex physical phenomena present in WR stars
%(Abbott \& Conti, 1987). 
	
%	In spite of the strong radiation fields of these hot,
%	luminous stars, dust grain condensation occurs in the winds of
%	late WC type stars.

WR stars exhibit IR excess, which has been attributed to 
free-free emission in the case of WN and early-WC type WR stars,
and thermal emission from circumstellar dust in the case of late-WC type
WR stars. 
In fact, several late-WC
stars have been discovered from infrared survey as these objects are highly 
obscured
in the optical by the circumstellar dust. Initially, the dust grains around
WC stars were thought
to be graphite (Cohen 1975), but it is widely established now that
the grains are made rather of
amorphous carbon (Williams et al. 1995; van der Hucht et al. 1996).
Episodic dust formation has been witnessed in a few WC7 stars 
from infrared photometric studies
and linked with the orbital motion of a suspected 
binary system (Williams et al. 1994).

The circumstellar environment around WR stars has been investigated so
far mainly by infrared photometry and spectroscopy.
Spatially resolved observations of   the circumstellar emission 
around this interesting class of sources in the optical and infrared
wavelengths are yet to be carried out on a regular basis, because of the  
rarity of these objects and of their distance.
High angular resolution studies of WR stars could improve our understanding
of high-mass stellar evolution and physical processes such as mass loss and
dust condensation.

Methods such as speckle or long baseline interferometry in the optical
and near infrared lack sufficient
angular resolution or sensitivity in order to resolve these sources. 
To our knowledge, only in one case
(namely Ve 2-45) has the circumstellar dust around one WR star 
been partially resolved 
by one-dimensional IR speckle observations (Allen, Barton \& Wallace 1981;
Dyck, Simon \& Wolstencroft 1984).
The results of these 
observations were
consistent with the dust shell model calculations. 
	
Lunar
occultations in the near-infrared have been very successful in
recent times in resolving circumstellar dust around evolved
late-type giants, supergiants and carbon stars 
(Richichi et al. 1998a;
Ragland, Chandrasekhar \& Ashok 1997).
%	The 10$\mu$m interferometry of
%	cool giants and supergiants have improved our understanding on
%	the circumstellar dust environment of these class of objects 
%	(Danchi et al., 1994).
In this paper, we report on lunar occultation observations of the
Wolf-Rayet star WR~112 (GL 2104), recorded almost 
simultaneously and independently from two telescopes in the near infrared.
The circumstellar dust shell is well resolved from these observations,
which for the first time have reached a resolution 
at the level of 0.003 arcsecond on such an object.

WR~112 has been classified as WC9 (Massey \& Conti 1983).
An alternative
classification of WC8 can also be found in the literature (Allen et al. 1977;
Cohen \& Kuhi 1976). WR~112 has been discovered from
the AFGL survey, while it  was overlooked in the visual because of its
faint magnitude of $v$=18.8 (note that $v$ is a narrow filter not
completely equivalent to the broad band V filter, for which no
photometry is available in the literature). By comparison with
our near-IR photometry, it is clear that WR~112 is subject to
%a strong reddening (A$_{\rm V} \approx$12).
a strong reddening (A$_v \approx$13).
The distance is estimated to be 1.2\,kpc,
with an uncertainty of about $\pm$20\%  (van der Hucht
et al. 1988).
%and
%the visual extinction is A$_v$=13.2 (van der Hucht et al., \cite
%{van88}).
\section{Observations and Data Reduction}\label{observations}
The lunar occultation observations reported here were
carried out with the 1.5m TIRGO and 1.23m Calar Alto
telescopes. At the time of observation, the primary
mirror of this latter was partially covered because of optical
imperfections and the effective size was only about 70\,cm.
The recorded event was a disappearance event at the
dark limb of the Moon.  Table~\ref{geo} lists the 
circumstances of the events. 
		
The instruments used were FIRT and FIRPO at TIRGO and Calar
Alto respectively. The instrumental details can be 
found in Richichi et al. (1998b) and references therein.
In both cases a standard K broad band filter was used to record the events.
		
In addition to lunar occultation observations, near infrared
photometry of this Wolf-Rayet star has been carried out at the
1.23m Calar Alto telescope one day after the lunar occultations,
and yielded the results
H$=6.37\pm0.1$ mag and K$=4.21\pm0.1$ mag.

The occultation data have been rebinned to improve the SNR.
This has been achieved at the
expense of a loss in angular resolution, which in any
case is largely sufficient to resolve the dust shell.
The resultant angular resolutions are 3 and 6 milliarcseconds (mas)
 respectively for the Calar Alto and
TIRGO observations. The rebinned light curves have been 
analyzed using a model independent algorithm
(CAL) introduced by Richichi (1989) for lunar occultation work. 
CAL is a composite algorithm
which makes use of non-linear least squares method and Lucy's
deconvolution algorithm,  wherein an arbitrary profile is assumed
as a initial guess and is iteratively
modified to fit the observed occultation data.
It is noteworthy that the solution to the problem of recovering
the brightness profile is not unique. The CAL algorithm
converges towards the most likely solution, which is not
necessarily the correct one. Simulations have shown that a
uniform disk will be recovered as a gaussian profile, 
and in Richichi (1989) a ratio
of 1.58 between the diameter of a uniform disk and the
FWHM of the equivalent gaussian was derived.
These considerations will be used in the discussion of our results.
	\begin{table}
	\begin{flushleft}
	\caption[]{Circumstances of the occultation events}
	\label{geo}
	\begin{tabular}{lrr}
	\hline
	\noalign{\smallskip}
	&\multicolumn{1}{c}{Calar Alto}&
	\multicolumn{1}{c}{TIRGO}\\
	\noalign{\smallskip}
	\hline
	\noalign{\smallskip}
	Date		&		14-08-97&	14-08-97\\
	Predicted Time (UT)&		22:36:54&	22:49:06\\	
	Predicted Position angle &	66$\degr$&	54$\degr$\\
	Predicted Contact angle  & 	$-$19$\degr$&	$-$31$\degr$\\
	Lunar phase 	&		84$\%$&		84$\%$\\
	Projected limb speed (km/s)&	0.680 &	 	0.693 \\
	\noalign{\smallskip}
	\hline
	\end{tabular}
	\end{flushleft}
	\end{table}

%	\begin{table}
%	\begin{flushleft}
%	\caption[]{Near infrared photometry of WR~112}
%	\label{phot}
%	\begin{tabular}{lrr}
%	\hline
%	\noalign{\smallskip}
%	\multicolumn{1}{c}{Filter Band}&
%	\multicolumn{1}{c}{H}&
%	\multicolumn{1}{c}{K}\\
%	\noalign{\smallskip}
%	\hline
%	\noalign{\smallskip}
%	Magnitude&$6.27 \pm 0.03$&$3.99 \pm 0.03$\\
%	\noalign{\smallskip}
%	\hline
%	\end{tabular}
%	\end{flushleft}
%	\end{table}

%\section{Observational history of WR~112}\label{history}
%WR~112 has been extensively studied by infrared spectroscopy.
%A medium resolution spectrum of this source taken with ISO clearly shows
%that the circumstellar grains are amorphous carbon in nature (van der
%Hucht et al. 1996). Dust formation in this source is believed to be
%steady at least for the last two decades from infrared photometry (van
%der Hucht et al. 1996).

\section{Results and Interpretation}\label{results}
\begin{figure*}
%\resizebox{\hsize}{!}{\includegraphics{fig1.ps}}
\centerline{\epsfysize=16cm \epsfbox{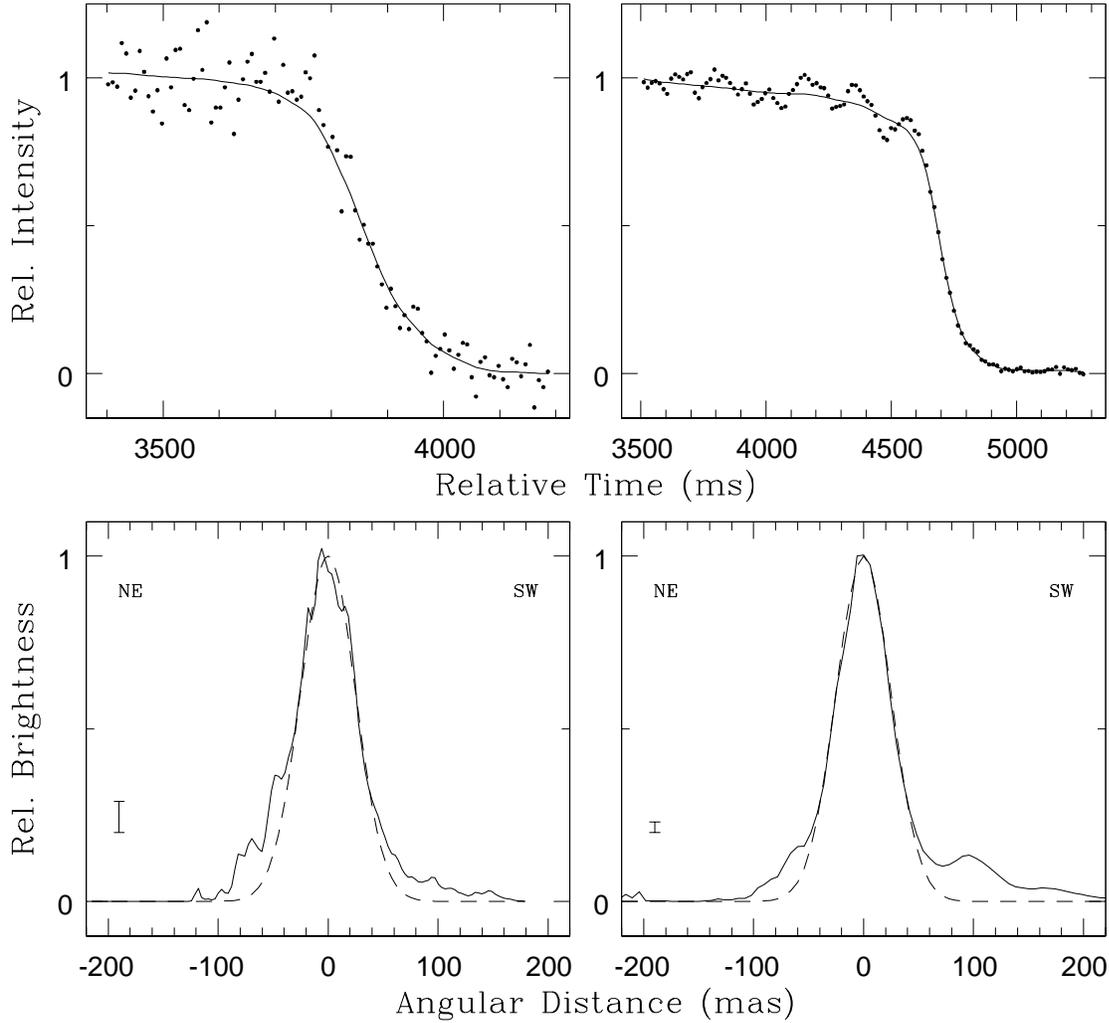} }
%\vskip 14cm
\caption{Upper panels:
occultation light curves of WR~112 (dots) observed at  Calar
Alto (left) and TIRGO (right) at 2.2$\mu$m. Also shown are the best fits
(solid lines). Lower panels: brightness profiles
recovered from the two data sets (solid lines). Also shown is a gaussian 
profile with FWHM of 59\,mas (dashed line), and the formal error bars
associated with the reconstruction in the two cases.
}
\label{fit}
\end{figure*}
%\begin{figure*}
%\resizebox{\hsize}{!}{\includegraphics{fig1.ps}}
%%\centerline{\epsfysize=7cm \epsfbox{bp.ps} }
%\caption{The recovered brightness profiles of WR~112 from our model fit
%to the observed light curves (continuous line). Also shown is a gaussian 
%profile with FWHM of 59\,mas (dotted line) superimposed on the brightness
%profile}
%%\label{bp}
%\end{figure*}

The observed light curves and the corresponding fits
are shown in the two upper panels of Fig.~\ref{fit},
while the recovered brightness profiles for each data set 
are shown in the lower panels of the same figure. Also shown are the formal
error bars on the recovered brightness profiles.
The scan direction for the two profiles differed by only 12$\degr$
due to the geometry of the events at the two sites
(see Table~\ref{geo}) and was roughly in a South-West $-$ North-East direction. 
This limits the possibility to exploit
significantly the potential for 2--D information, but it has on the
other hand the advantage that the analysis of the
two observations can be independently checked,
thus adding reliability to the result. 
We attribute the extended brightness profile of WR~112 to the presence of 
warm circumstellar dust close to the stellar photosphere,
as discussed below.
The data from Calar Alto were noisier because of
the smaller telescope aperture,  and this is reflected in a
lower SNR in the reconstructed profile. Even so, the
general features are consistent with the profile obtained
from the TIRGO data. 
%We will then use this latter, higher SNR profile
%for the discussion.

We note the following points:
{\em a)} 
there is no evidence of the central star, which is expected to be
completely unresolved and should leave a distinctive signature (well defined
diffraction fringes) in
the occultation trace;
{\em b)} 
the profile of this dust shell can be approximately fitted
by a gaussian with a FWHM of $0\farcs059$, or $\approx71$\,AU;
{\em c)} 
the profile has broader wings than
the simple gaussian approximation. 

These wings are not symmetric, with
the one in the 
%North-East direction being sharper and that in
South-West direction significantly broader. 
This is made more evident by the comparison
with the gaussian profile, which is also shown in Fig.~\ref{fit}.
It can be noted that the recovered broad feature 
to the SW is well above the
noise level of the TIRGO data, and is present also in
the lower SNR data of Calar Alto. The feature to the
NE is less evident, but is also recovered consistently in both data sets.

A two component blackbody fit to the photometric data suggests that the 
central star is about 6 magnitudes fainter than the circumstellar shell 
in the K band. This is well below our detection limit, and
explains why there is no evidence of the central star from our observations.

The peculiar continuum and the weakness of the CIII
line at 4650\AA\, in the optical spectra suggest a possible companion
of early spectral type (Cohen \& Kuhi 1976;
Cohen \& Vogel 1978; Massey
\& Conti 1983): detection of the companion is hampered by the same factors as
the WC star, namely being about 6 magnitudes fainter than the dust shell at 2
microns.
%proposed early type companion should have about the same
%brightness as that of the primary in the K band:
%the limited SNR in our light curves rules out the possibility to detect it.

One important aspect in the interpretation of the data is the
fact that, in spite of the strong reddening, the dust around WR~112 
is optically thin. This is deduced from the fact that
the IR excess (which has been derived from the model fit to the observed 
spectral 
energy distribution) 
amounts to only 7.9\% of the total flux 
(Williams, van der Hucht, Th\'{e} 1987).
Most of the reddening must be interstellar in origin, and this
is consistent with the fact that
WR~112 lies in a region of high interstellar extinction
(l$_{\rm II}$=12$\degr$,
b$_{\rm II}$=$-1\degr$). Moreover, it has been shown that
%A$_{\rm v}$ can be well correlated with the strength of
A$_v$ can be well correlated with the strength of
interstellar absorption features at
3.4\,$\mu$m (Standford, Pendleton \& Allamandola 1995)
%Imanishi et al. 1996).
and 9.7\,$\mu$m (Roche \& Aitken 1984).
In the case of
WR~112 this leads to the conclusion that circumstellar extinction 
contributes very little to the
observed reddening.
%value A$_{\rm V}\approx$10 which is
%well in agreement with the observed reddening.
Also, there is strong indirect evidence from the photometric
monitoring of WR~112 in the infrared and from its spectral type, that the dust
condensation is probably a continuous process in this star.
WR~112 has been searched for HCN molecular line emission with
negative results (Clair et al. 1979).
Recently, the detection of the radio continuum at 3\,cm
has been reported by Leitherer et al. (1997).
There is no appreciable polarization in this source from the infrared
measurements carried out by  Kobayashi et al. (1978).

The FWHM of the recovered profile, when fitted with a gaussian,
is 59$\pm$3\,mas. The corresponding uniform disk
angular diameter is 93$\pm$5\,mas, from the arguments 
given in Sect.~\ref{observations}.
Allen et al. (1981) gave a scaling factor of 1.5
between the radii of a uniform disk and of the inner dust shell radius, 
under the assumption of constant dust condensation and optically thin shell
which we have just discussed.  Thus we derive an inner dust shell angular radius 
of 31\,mas. At an assumed distance of 1.2\,kpc to the source and including the 
associated uncertainty, this
value translates to (5.6 $\pm$ 1.2) $\times$ 10$^{14}$ cm,
or $\approx 37$\,AU.
% and is subject to have an uncertainty of about 20$\%$.
By comparison, Allen et al. (1981) derived a value of 
$1.2 \times 10^{15}$\,cm for 
the inner radius of the dust shell around another WC9 star, namely 
Ve 2-45, at 2.2$\mu$m from speckle observations. 

We also note that under the above mentioned assumptions,
the models 
predict a bright rim at the inner zone of the shell where dust condenses 
(Rowan-Robinson 1980;
Rowan-Robinson $\&$ Harris 1982). This is expected to be more
sharply defined at 2$\mu$m than in the visual or at longer wavelengths. 
Such a bright rim is not evident from our data, but we must caution
against a negative conclusion in this sense. In fact, we remind that
the profiles shown in Fig.~\ref{fit} are integrated in one direction
(the direction perpendicular to the lunar limb), and it can be
shown that in the case of circular symmetry the bright rim feature
is effectively canceled (see for instance Ridgway et al. 1986, 1987).

%We don't see
%any such bright-rim from both our observations carried out at 2.2$\mu$m. 
%Such sharp bright-rim
%(spike like 
%feature), if any, present 
%should have resulted in
%diffraction fringes as in the case of binary stars 
%(Richichi et al., \cite {Richichi97}) which
%should have been easily detected from 
%our observations carried out with adequate angular resolution and
%sensitivity.
%	
Adopting the chemical composition of the grains to be amorphous carbon,
the equilibrium grain temperature is expected to fall off with the distance
from the star as r$^{-0.4}$ (Williams et al. 1994).
Assuming the stellar 
effective temperature to be 19000\,K and the stellar 
radius to
be 14.6\,R$_\odot$ 
%(van der Hucht et al. 1986), 
(van der Hucht, Cassinelli \& Williams 1986),
we derive the grain temperature at the inner dust shell to be
1150 $\pm$ 270\,K. The inner dust shell radius is thus at 550 $\pm$ 160\,R$_\star$. 
The derived dust grain temperature is consistent with the expected dust
temperature at the condensation zone. In Table~\ref{parameters}, we summarize 
the adopted
stellar parameters and the dust grain characteristics, and the derived dust shell
parameters.

\begin{table}
\begin{flushleft}
\caption[]{Stellar and dust shell parameters}
\label{parameters}
\begin{tabular}{ll}
\hline
\noalign{\smallskip}
%Parameters&Value& Remarks\\
%\noalign{\smallskip}
%\hline
\noalign{\smallskip}
Stellar effective temperature           & 19000\,K\\
%        &Ref: van der Hucht et al. (1986)\\
Stellar radius                          & 14.6\,R$_\odot$\\
Distance		                & 1.2\,kpc\\
Dust grain composition                  & amorphous carbon\\
Dust grain temperature                  & T$_{\rm grain}\propto$ r$^{-0.4}$\\
Dust grain density                      & $\rho_{\rm grain}\propto$ r$^{-2}$\\
\hline
Inner dust shell angular radius         & $0\farcs031$\\
Inner dust shell linear radius          & 5.6$\times$10$^{14}$\,cm $\approx$ 550\,R$_\star$\\
% & this paper\\
Inner dust shell temperature            &1150\,K\\
\noalign{\smallskip}
\hline
\end{tabular}
%$^1$r denotes the radial distance 
%of the dust grains
%from the central star
%\end{tabular}

\end{flushleft}
\end{table}

%However, our accurate measurement on the angular size of the
%dust shell at 2.2$\mu$m will application when the knowledge
%of effective temperature and radius improves.
We compare our derived dust shell parameters with those estimated by Williams et al. (1987) 
from their detailed dust shell model fits to the observed photometric data.
They estimate
the dust grain temperature at the inner dust zone 
and the inner dust shell radius to be 960\,K and 690\,R$_\star$ 
respectively.
Our derived dust shell parameters are
in good agreement with their estimation.
More recently, Zubko (1998) estimate the dust grain temperature at the inner dust zone
of 1085\,K and the inner dust shell radius of 1500\,R$_\star$ for this WR star. 
While their estimated dust
temperature is consistent with our observations, their value for the dust shell inner radius 
is about a factor of 2.5 larger than our determination.
However, this author has assumed different values for the 
stellar photospheric parameters (T$_{\rm eff}$ =22000\,K and R$_\star$ = 10\,R$_\odot$).
We also note that 
the emphasis is more on far-infrared fluxes than near-infrared measurements (J and H band): 
on the other hand,
near-infrared fluxes are probably more sensitive to the 
inner dust shell radius. 

It is quite probable that the stellar radii of WC9 stars may be much smaller 
than the 
value adopted here, ie. R$_\star$ = 3\,R$_\odot$, and a correspondingly larger
effective temperature of T$_{eff}$ = 35000\,K, but without much change in the
stellar luminosity (Crowther, 1997). In this case, the
dust shell size reported here would not change in absolute units, but the inner
dust shell radius would be thus located at 2680 $\pm$ 800\,R$_\star$ at 
the grain temperature of 1130\,K.
	
The suspected hot companion of WR~112, if present, would also heat up the
circumstellar dust shell: the combined luminosity should be then adopted in
estimating the dust grain temperature at the inner dust shell zone 
(Williams, van der Hucht, Th\'{e} 1987). In this case, the derived grain
temperature at the inner shell radius would be larger by about 18$\%$ compared
to the case when no companion is assumed.

\section{Conclusions}\label{conclusions}
We have investigated the circumstellar dust environment around a
the WC9 Wolf-Rayet star 
WR~112 from near-infrared lunar occultation observations. The circumstellar
dust shell has been directly detected for the first time from these
observations carried out at 2.2$\mu$m from two independent telescopes. 

The recovered brightness profile has a
FWHM of $0\farcs059$, or about 71\,AU. Under reasonable assumptions of
small optical thickness of the dust and constant dust
formation, we also derive some physical parameters for the
dust shell: inner radius of dust formation $0\farcs031$ or $\approx$37\,AU,
and temperature of condensation $\approx 1150$\,K.
%We do not detect the bright rim predicted by
%model calculations, but this is not conclusive as the
%1--D integration intrinsic in lunar occultation observations
%effectively does not permit  such a detection.
%According to model predictions, a bright rim should be present 
%at the location of the dust condensation zone; this is not
%immediately apparent from our data, but this is not conclusive
%since the 1--D integration intrinsic in lunar occultation
%observations would not permit us to reveal such a feature.

Our observations have also shown departure from
circular symmetry, with wings in the brightness profile which
are more prominent in the South-West direction than in the
North-East one. The SW wing can be traced out to
about $0\farcs2$.
There is no signature of the central star from our observations,
and we show that this is consistent with the expected properties
of the object and the associated interstellar extinction.

The quantitative conclusions that we derive 
are in good agreement with theoretical predictions and
show the potential of high angular resolution observations
to investigate the inner structure of this class of objects.
Further similar observations would be very useful to constrain
further the physical characteristics of this object, and to
study the details of its structure. The current series of 
lunar occultations of WR~112 will continue to the end of 1999 and will 
be visible for all equatorial and southern latitude observatories.
%(ADD: next occultations?)

\section*{Acknowledgments}
%\acknowledgements
The authors would like to thank the referee Prof. Peredur Williams for his 
valuable comments which have greatly improved the quality of this paper. 
We also thank Prof. Mario Perinotto for his useful suggestions.
This research has made use of the {\em Simbad} database, operated at
CDS, Strasbourg (France).


\begin{thebibliography}{99}
%\begin{thebibliography}{}

\bibitem[1987]{Abbott87}
Abbott D.C., Conti P.S., 1987, ARA\&A, 25, 113
\bibitem[1981]{Allen81}
Allen D.A., Barton J.R., Wallace P.T., 1981, MNRAS, 196, 797
\bibitem[1977]{Allen77}
Allen  D.A.,  Hyland A.R.,  Longmore A.J.,  Caswell
J.L., Goss W.M., Haynes R.F.,
1977, ApJ, 217, 108
%\bibitem{bh}
%Barlow \& Hummer, 1982, 
\bibitem[1975]{Cohen75}
Cohen M., 1975, A\&A, 40, 291
\bibitem[1976]{Cohen76}
Cohen M, Kuhi L.V., 1976, PASP, 88, 535
\bibitem[1978]{Cohen78}
Cohen M., Vogel S.N., 1978, MNRAS, 185, 47
\bibitem[1979]{Clair79}
Clair D.A.,  Dickinson D.F.,  Gottlieb C.A.,
Gottlieb E.W.,
1979, PASP, 91, 830
\bibitem[1997]{Crowther}
Crowther P.A., 1997, MNRAS, 290, L59
%\bibitem[1994]{Danchi}
%Danchi, W.C., Bester, M., Degiacomi, C.G., Greenhill, L., Townes, C.H.,
%1994, AJ, 107, 1469
\bibitem[1984]{Dyck84}
Dyck H.M., Simon T., Wolstencroft R.D., 1984, ApJ.,
277, 675
%\bibitem[1996]{Imanishi96}
%Imanishi M., Sasaki Y., Goto M., Kobayashi N., Nagata T., Jones T.J.,
%1996, AJ, 112, 235
\bibitem[1986]{van86}
van der Hucht K.A., Cassinelli J.P., Williams P.M., 
1986, A\&A, 168, 111
\bibitem[1987]{van87}
van der Hucht K.A., Williams P.M., Th\'{e} P.S.,
1987, QJRAS, 28, 254.
\bibitem[1988]{van88}
van der Hucht, K.A.,  Hidayat, B., Admiranto, A.G., Supelli, K.R. \&
Doom, C., 1988, A\&A, 199, 217
\bibitem[1996]{van96}
van der Hucht K.A.,  Morris P.W.,  Williams P.M. et al.,
1996, A\&A, 315L, 193
\bibitem[1978]{Kobay78}
Kobayashi Y., Kawara K., Maihara T., Okuda H., Sato S.,
Noguchi K.,
1978, PASJ, 30, 377 
%\bibitem[1991]{Lang91}
%Lang K.R., 1991 in 'Astrophysical data : Planets and stars', Springer
%publishers, Berlin
\bibitem[1997]{Lei97}
Leitherer C.,  Chapman J.M., Koribalski B., 1997, ApJ, 481, 898
\bibitem[1983]{MC83}
Massey P., Conti P.S.,
1983, PASP, 95, 440
\bibitem[1997]{Ragland97}
Ragland S., Chandrasekhar T., Ashok N.M., 1997, A\&A, 319, 260
\bibitem[1984]{RA84}
Roche P.F., Aitken D.K., 1984, MNRAS, 208,481
\bibitem[1980]{RR80}
Rowan-Robinson M., 1980, ApJS, 44, 403
\bibitem[1982]{RRH82}
Rowan-Robinson M., Harris S., 1982, MNRAS, 200, 197
\bibitem[1989]{Richichi89}
Richichi A., 1989, A\&A 226, 366
%\bibitem[1998]{Richichi_III}
%Richichi, A., Ragland, S., Fabbroni, L., 1998b, A\&A, 330, 578 
\bibitem[1998b]{Richichi98b}
Richichi A., Ragland S., Stecklum B., Leinert Ch., 1998b, A\&A, 
(in press)
%\bibitem[1997]{Richichi97}
%Richichi A., Calamai G., Leinert Ch., Stecklum, 1997, A\&A, 322, 202
\bibitem[1998a]{Richichi98a}
Richichi A., Stecklum B., Herbst T.M.,
Lagage P.-O., Thamm E., 1998a, A\&A, 334, 585
\bibitem[1986]{Ri86}
Ridgway S.T., Joyce R.R., Connors D., Pipher J.L., Dainty C., 1986, 
ApJ, 302, 662
\bibitem[1987]{Ri87}
Ridgway S.T., Joyce R.R., Connors D., Pipher J.L., Dainty C., 1987, 
ApJ, 312, 963 (Erratum)
\bibitem[1995]{S95}
Standford S.A., Pendleton Y.J., Allamandola L.J., 1995, ApJ, 440, 697
\bibitem[1987]{W87}
Williams P.M., van der Hucht K.A., Th\'{e} P.S.,
1987, A\&A, 182, 91.
\bibitem[1995]{W95}
Williams P.M., Cohen M., van der Hucht K.A., Bouchet P.,
Vacca W.D., 1995, MNRAS, 275, 889
\bibitem[1994]{W94}
Williams P.M.,  van der Hucht K.A.,  Kidger M.R.,
Geballe T.R., Bouchet P.,
1994, MNRAS, 266, 247
\bibitem[1998]{Zubko98}
Zubko V.G., 1998, MNRAS, 295, 109
\end{thebibliography}
\end{document}